\documentclass[twocolumn,showpacs,prl,preprintnumbers,amsmath,amssymb]{revtex4}

\usepackage{graphicx}
\usepackage{dcolumn}
\usepackage{bm}
\usepackage{appendix}
\usepackage{hyperref}

\begin{document}

\title{Protected percolation: a new universality class pertaining to heavily-doped quantum critical systems}
\author{Sean Fayfar$^{1}$, Alex Breta\~{n}a$^1$, Wouter Montfrooij$^{1}$}
\affiliation{$^1$Department of Physics and Astronomy, University of Missouri, Columbia, MO 65211, United States}

\begin{abstract}
    We present the results of computer simulations on a class of percolative systems that forms a new universality class. We show the results for the critical exponents for this new class, inferred from simulations of two- and three-dimensional lattices consisting of up to one billion lattice sites. These new percolative systems differ from standard percolative systems in that once a cluster breaks off the lattice spanning cluster, its sites become protected and cannot be removed. This situation closely mimics the situation in heavily-doped quantum critical systems where isolated magnetic clusters are protected from (further) Kondo screening. Our results indicate that protected percolation violates the Harris criterion, which yields a natural explanation as to why universal exponents for quantum phase transitions have been elusive.
\end{abstract}

\date{\today}

\maketitle

\section{}
Percolation is a phenomenon that applies to many physical systems whose critical behavior falls into known universality classes \cite{Stauffer1985,Nakayama1994,Sahini1994}. Percolation describes how a system responds to the removal of its elements or the connections between them. When enough of the connections are broken, the percolation threshold - the point where the system spanning connection fractures - is reached; the dimensionality and connectivity of the system determines when this happens. As a system approaches this threshold, universal behavior is displayed: behavior characterized by critical exponents that only depend on the dimensionality of the system and ordering quantity.   

We describe a new type of percolation - inspired by observations on heavily-doped quantum critical systems \cite{Montfrooij2019} – whose critical behavior falls in a new universality class. This new percolative system, dubbed protected percolation, has the restriction that upon emptying a lattice only sites attached to the lattice spanning cluster can be removed. The isolated clusters that form become ``protected" from further removal. This procedure is inspired by heavily-doped quantum critical systems that exhibit a magnetic lattice whose moments are effectively removed through Kondo shielding as the system is cooled \cite{Sachdev2011}. However, when magnetic clusters break off the lattice spanning cluster, the moments of the magnetic ions align with their neighbors because of quantum mechanical finite-size effects \cite{Heitmann2014};  since ordered moments are not likely to be Kondo shielded because of the spin-flip interaction at the heart of this shielding\cite{Kondo1964,Montfrooij2019}, such clusters are protected from further degradation. Therefore, this new percolation model is relevant to (heavily-doped) quantum critical systems. 

The three-dimensional manifestation of protected percolation violates the Harris criterion \cite{Harris1974} whereas the two-dimensional manifestation satisfies this criterion. The Harris criterion states that impurities do not alter how a system transitions from a disordered to an ordered phase. This critical behavior is captured by a universal set of critical exponents that describe how, for instance, the order parameter  (exponent $\beta$) or the mean cluster size (exponent $\gamma$) depend on the distance to the critical point. When the Harris criterion is satisfied,
\begin{equation}
    \gamma+2\beta>2,
    \label{eq:harrisCriterion}
\end{equation} 
then impurities can shift the critical point, but they do not alter the values of the critical exponents \cite{Harris1974}; when the criterion is violated ($\gamma+2\beta<2$), then the critical exponents become system dependent and universal behavior disappears. We show that in protected percolation the criterion is violated for three-dimensional lattices, but it is satisfied for two-dimensional lattices. 

The outline of this paper is as follows. We first discuss our Monte Carlo (MC) computer simulations we performed to obtain the critical exponents. Next we show the results for various lattice types and demonstrate that protected percolation falls in a new universality class. We end with a discussion on the Harris criterion and its implications for the low-temperature behavior of strongly correlated electron systems that have been doped to be at the quantum critical point.

\begin{figure}[h]
    \includegraphics[width=\columnwidth]{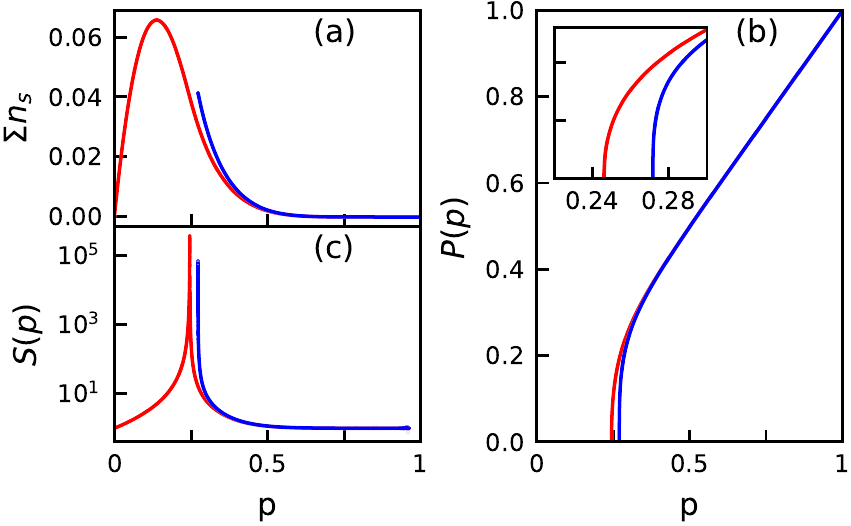}
    \caption{(Color online) The results of MC simulations for a body-centered cubic lattice of size $1000^3$ averaged over 100 iterations. Standard and protected percolation are shown in light gray (red) and dark gray (blue), respectively. (a) The zeroth moment (Eqn. \ref{eq:clusmoments} with $k=0$), being the mean cluster number. (b) The strength of the lattice spanning cluster (Eqn. \ref{eq:strPercClus}). (c) The average size of the clusters (Eqn. \ref{eq:clusSize}).}
    \label{fig:clusMoments}
\end{figure}

In percolation theory \cite{Stauffer1985}, the point where (upon emptying a lattice) the lattice spanning connection fractures is called the percolation threshold, denoted by $p_c$. The fractional occupation of the lattice is denoted by $p$ (0 $\leq p \leq$ 1) . This threshold depends on the dimensionality of the lattice and the number of neighbors. As the system approaches the percolation threshold, both the strength of the lattice spanning cluster , $P(p)$, change exponentially as do some of the moments of the cluster size distribution, $M_k(p)$. We define the k$^{th}$ moment  of the cluster size distribution as
\begin{equation}
    M_k(p) = \sum_{s}s^k n_s(p),
    \label{eq:clusmoments}
\end{equation}
where $s$ is the number of sites in a cluster and $n_s$ is the number of clusters per lattice site containing s sites. The zeroth moment ($k=0$) represents the mean number of clusters per lattice site and its second derivative shows non-analytical critical behavior \cite{Stauffer1985}. The first moment ($k=1$) represents the probability that an arbitrary site belongs to an isolated cluster, and it relates to $P(p)$ which is defined by:
\begin{equation}
    P(p) = p - \sum_{s} sn_s(p).
    \label{eq:strPercClus}
\end{equation}
The strength of the lattice spanning cluster exhibits critical behavior near the percolation threshold. Finally, the second moment ($k=2$) is used to calculate the mean cluster size, $S(p)$, as
\begin{equation}
    S(p) = \frac{\sum_s s^2 n_s(p)}{\sum_s s n_s(p)} \approx \frac{\sum_s s^2 n_s(p)}{p_c},
    \label{eq:clusSize}
\end{equation}
where we have used that $S(p)$ becomes approximately equal to the second moment close to the percolation threshold since $P(p_c)=0$. The results of simulations for Eqns. \ref{eq:clusmoments}~--~\ref{eq:clusSize} are shown in Fig.~\ref{fig:clusMoments}.

The strength of the lattice spanning cluster and the second moment of the cluster size distribution display power law dependences when measured as a function of how far the system is from the percolation threshold $p-p_c$ \cite{Stauffer1985}. The distribution of cluster numbers at the percolation threshold $n_s(p_c)$ also shows a power law dependence \cite{Stauffer1985}. All this is expressed as
\begin{align}
    P(p) &= P_0 (p-p_c)^\beta + (p-p_c)
    \label{eq:strpercclusPowerLaw}\\
    S(p) &= S_0 (p-p_c)^{-\gamma}
    \label{eq:clussizePowerLaw}\\
    n_s(p_c) &\sim s^{-\tau}
    \label{eq:tauPowerLaw}
\end{align}
where $\beta$, $\gamma$, and $\tau$ are the critical exponents. We will distinguish between the critical exponents for standard and protected percolation by putting a prime on the latter. 

Protected percolation requires a different MC algorithm than standard percolation. In particular, it is no longer possible to simulate a lattice at a particular occupancy, rather the lattice needs to be emptied one site at a time. In standard percolation a single instance of a lattice is randomly generated for a given occupancy after which the lattice is analyzed to determine its connectivity. However, in protected percolation sites cannot be removed from isolated clusters and hence, the prior history of when clusters become isolated needs to be taken into account. As such, protected percolation requires a lattice to be emptied step by step rather than to be analyzed at a single occupation. This greatly increases the computational overhead for carrying out the MC computer simulations. 

In order to get around the computational overhead of determining cluster connectivity at every occupancy, we opted for filling the lattice one site at a time and use bookkeeping to correct for forbidden site removal from isolated clusters. Upon filling, it is easy to check whether a newly occupied site attaches to an existing cluster, connects existing clusters, or forms a new isolated cluster. We have opted to use a disjoint-set data structure \cite{Galler1964} developed by Newman and Ziff \cite{Newman2000,*Newman2001} to analyze the connectivity of the lattice. Bookkeeping is needed upon filling because some events that are included are forbidden in protected percolation, such as the merging of two isolated clusters as this would correspond to the breaking up of an isolated cluster upon emptying. We must retroactively correct  our data as a function of occupancy. Whenever such a forbidden event happens, we add this to the calculated shift in threshold between standard and protected percolation. We verified this procedure by comparing the percolation threshold and the moments of the cluster distribution for a particular set of random numbers using our method with a Hoshen-Kopelman algorithm \cite{Hoshen1976} while emptying a lattice and found identical results. \footnote{More than one lattice spanning cluster may appear and we have chosen to only use simulations containing a single percolating cluster for Eqn. \ref{eq:clusmoments}~--~\ref{eq:clusSize}.} 

\begin{table}[t]
    \caption{\label{tab:pc} Percolation thresholds for both standard and protected percolation as determined from our simulations.}
    \begin{ruledtabular}
        \begin{tabular}{llll}
            \textrm{Lattice}&
            \textrm{Literature $p_c$}&
            \textrm{Standard $p_c$}&
            \textrm{Protected $p_c$}\\
            \colrule
            Square & 0.59274601(2)~\cite{Jacobsen2014} & 0.592758(18) & 0.602752(11)\\
            Triangular & 0.5~\cite{Stauffer1985} & 0.500009(20) & 0.508691(13)\\
            Honeycomb & 0.697040230(5)~\cite{Jacobsen2014} & 0.697056(28) & 0.707168(17)\\
            \hline
            SC & 0.3116081(13)~\cite{Ballesteros1999} & 0.311593(7) & 0.3423156(14)\\
            BCC & 0.2459615(10)~\cite{Lorenz1998b} & 0.245966(7) & 0.2713933(15)\\
            FCC & 0.1992365(10)~\cite{Lorenz1998b} & 0.199239(4) & 0.2182667(11)\\
        \end{tabular}
    \end{ruledtabular}
\end{table}

\begin{figure}[h]
    \includegraphics[width=\columnwidth]{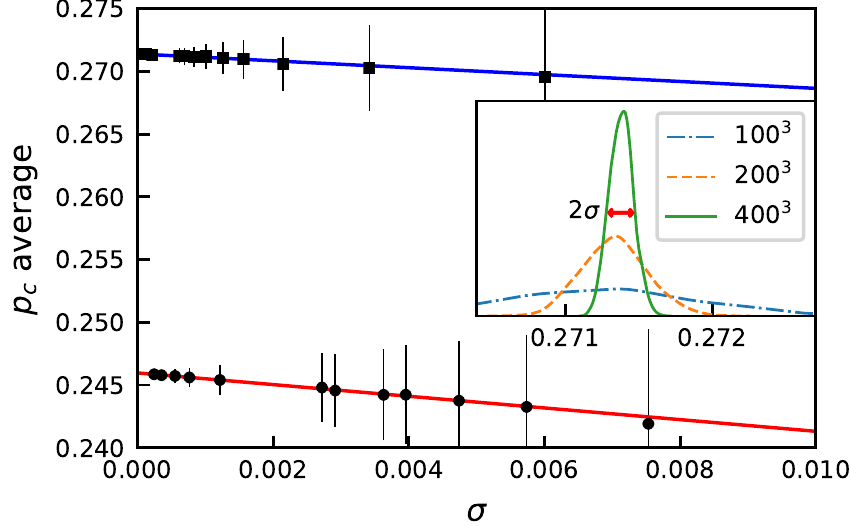}
    \caption{The inset shows the distribution of thresholds for protected percolation from 1000 iterations for three lattice sizes. The distribution is characterized by $p_{c}$ average and the standard deviation $\sigma$. With increasing system size, the distribution becomes more narrow while approaching the threshold for an infinite lattice. Using the relationship between $p_{c}$ average and $\sigma$, we obtain $p_c$ by extrapolating to $\sigma\rightarrow 0$. The fits for standard and protected percolation are shown by the lower (red) and upper (blue) lines, respectively. The error bars are given by $\sigma$}
    \label{fig:pcavg}
\end{figure}

We show our results for the percolation thresholds for various lattice types in Fig.~\ref{fig:pcavg} and Table~\ref{tab:pc}. We used an identical set of random numbers for both protected and standard percolation. Our results for standard percolation match those in the literature; we used the Levinshtein method \cite{Levinshtein1975} to calculate the thresholds. We ran separate simulations in order to determine the percolation threshold for 1000 iterations of varying lattice sizes (ranging from $40^3$ to $800^3$ sites).  

\begin{figure}[h]
    \includegraphics[width=\columnwidth]{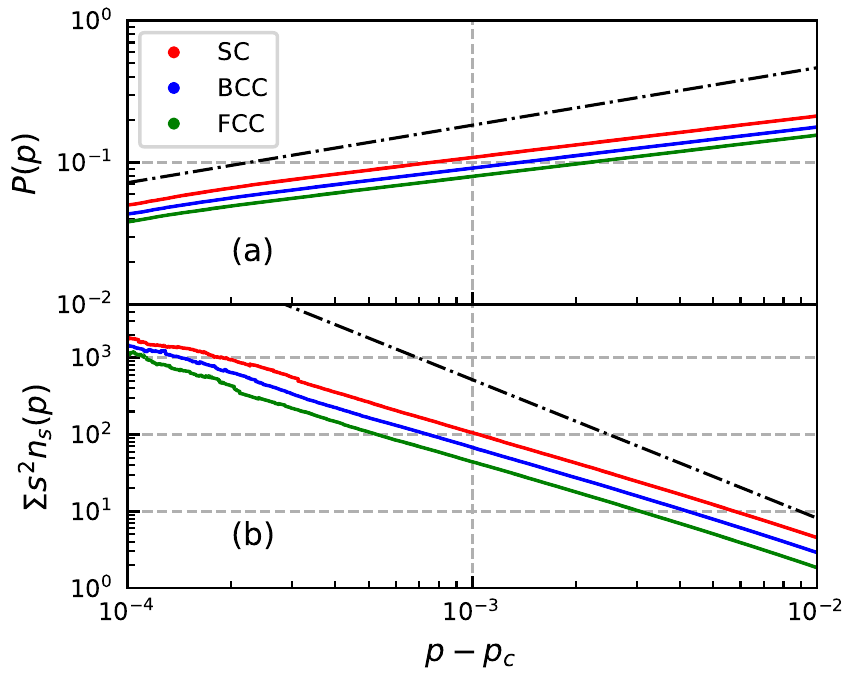}
    \caption{\label{fig:universality} Simulation results for 1000 iterations for 3D lattices of size $400^3$. Lattice types include SC (top), BCC (middle), and FCC (bottom). (a) The strength of the lattice spanning cluster. (b) The second moment of the cluster size distribution. In both cases, the lines appear parallel demonstrating the universality of protected percolation. The dashed-dotted lines represent the literature values for the slopes in standard percolation.}
\end{figure}

Using the extrapolated thresholds shown in Table \ref{tab:pc}, we plot the critical behavior of the second moment and $P(p)$ for protected percolation in Fig.~\ref{fig:universality} for simple cubic (SC), body-centered cubic (BCC), and face-centered cubic (FCC) lattices. Since each of the data sets have identical slopes we conclude that protected percolation displays universal critical behavior. Given that the slopes are substantially different from standard percolation, we conclude that protected percolation represents a new universality class (see Table \ref{tab:critexp}). 

There exists an analytical relation between the critical exponents of standard and protected percolation \cite{Heitmann2014}. In both scenarios, the strength of the lattice spanning cluster follows the same trajectory as sites are removed. When the percolation threshold is reached, the lattice spanning cluster will be identical in both cases. In the standard case, a site will be removed from the lattice spanning cluster (on average) every $p/P(p)$ removal steps. In the protected case, only sites from the lattice spanning cluster can be removed, so the strength of the lattice spanning cluster decreases at each step in $p$. As such, the lattice spanning cluster will undergo the same site removals in both standard and protected cases, but at a different value of $p$. Thus, we will always be able to find a solution to the following equation for any value of $p$ with $p>p_c$ :
\begin{equation}
    P(p)=P'(p').
    \label{eq:strpercEquality}
\end{equation}

Since the equation holds, we can then equate the slope of $P(p)$ to the slope of $P'(p')$ using the fact that the slope for the protected case will be steeper by a factor of $p/P(p)$ than the standard case (with $x=p-p_c$ and $x'=p'-p_c'$):
\begin{equation}
    \frac{dP'(x')}{dx'} = \frac{dP(x)}{dx}\frac{p}{P(x)}.
    \label{eq:derivative}
\end{equation}
Using Eqn.~\ref{eq:strpercclusPowerLaw} to describe critical behavior, we rewrite Eqn.~\ref{eq:derivative} as
\begin{equation}
    \beta' P_0' x'^{\beta'-1} = \beta P_0 x^{\beta-1} \frac{p}{P_0 x^\beta} = \frac{\beta (x + p_c)}{x} \approx \frac{\beta p_c}{x},
    \label{eq:derivApplied}
\end{equation}
where we have left out $(p-p_c)$ and $(p'-p_c')$ terms close to the percolation threshold because the power law term dominates. The solution to Eqn.~\ref{eq:strpercEquality} is found in a similar manner, yielding
\begin{equation}
    x = \left( \frac{P_0'}{P_0} \right) ^{1/\beta} \left( x' \right) ^{\beta' / \beta}.
    \label{eq:xsolution1}
\end{equation}
Substituting Eqn.~\ref{eq:xsolution1} into Eqn.~\ref{eq:derivApplied}, we find  \cite{Heitmann2014}
\begin{equation}
    \beta' P_0' \left(x'\right)^{\beta' - 1} = 
    \frac{\beta p_c}{x} = 
    \frac{\beta p_c}{ \left( \frac{P_0'}{P_0} \right)^{1/\beta} \left( x' \right) ^{\beta'/\beta} }.
    \label{eq:subeqns}
\end{equation}

\begin{figure*}[]
    \includegraphics[width=\textwidth,clip]{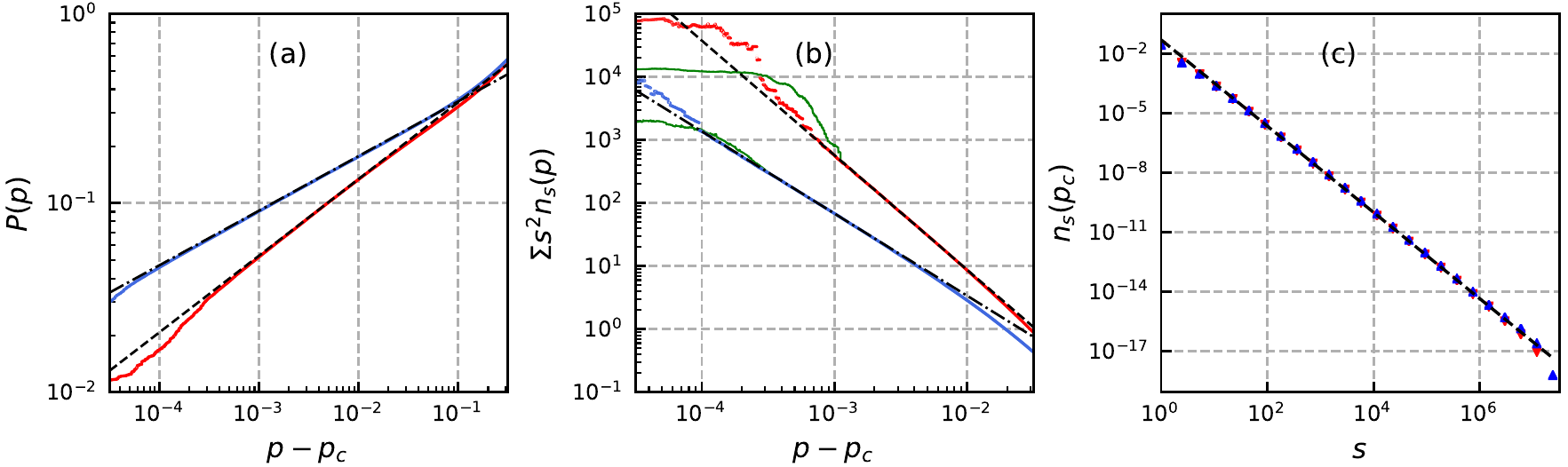}
    \caption{The results of simulations of a BCC lattice of size $1000^3$ averaged over 100 iterations. The fitted exponents are given in Table \ref{tab:critexp}. (a) Log-log plot of $P(p)$ with fits given by Eqn.~\ref{eq:strpercclusPowerLaw} with standard (lower) and protected (upper). (b) Log-log plot of the second moment with fits given by Eqn.~\ref{eq:clussizePowerLaw} for standard (upper) and protected (lower). Included in green are the results from simulations of a $400^3$ lattice averaged over 1000 iterations. (c) Log-log plot of the cluster numbers at the percolation threshold with a fit given by Eqn.~\ref{eq:tauPowerLaw} wfor standard (inverted triangle) and protected (triangle) simulation results.} 
    \label{fig:critExpFits}
\end{figure*}

Since this equation holds for all values of $x'$, we must have that the prefactors as well as the powers of $x'$ match on both sides. Solving this, we find
\begin{align}
    \label{eq:betaRelation}
    \beta' &= \frac{\beta}{1 + \beta}\\
    P_0' &= \left( P_0 \right) ^{1/(\beta+1)} \left( (1 + \beta) p_c  \right) ^{\beta/(\beta+1)}.
    \label{eq:prefactorRelation}
\end{align}
These equations readily capture that protected percolation is a new universality class as $\beta' \neq \beta$.

Fitting the data shown in Fig.~\ref{fig:critExpFits} for the strength of the lattice spanning cluster, the second moment, and the cluster numbers at the percolation threshold for standard and protected percolation to Eqns. \ref{eq:strpercclusPowerLaw}, \ref{eq:clussizePowerLaw}, and \ref{eq:tauPowerLaw}, respectively, we obtain satisfactory fits (dashed lines in Fig.~\ref{fig:critExpFits}). The critical exponents obtained from the fits are given in Table~\ref{tab:critexp}.

\begin{table}[b]
    \caption{The critical exponents for both standard and protected percolation as determined from fits. The value for the Harris criterion in each case is also included. All 2D critical exponents are exact values provided scaling relations hold.}
    \label{tab:critexp}
    \begin{ruledtabular}
        \begin{tabular}{llll}
            \textrm{Critical exponent} & 2D & 3D Fits & 3D Literature\\
            \colrule
            \textrm{$\beta$} & 5/36 & 0.4053(5)  & 0.405(25)~\cite{Adler1990}\\
            \textrm{$\gamma$} & 43/18 & 1.819(3) & 1.805(20)~\cite{Adler1990} \\ 
            \textrm{$\tau$} & 187/91 & 2.1753(11) & 2.189(2)~\cite{Lorenz1998a} \\
            \textrm{$\gamma+2\beta$} $(= $d$\nu)$ & 2.667 & 2.6296(32) & 2.62(5) \\
            \hline
            \textrm{$\beta'$} & 5/41 & 0.28871(15) & \\
            \textrm{$\gamma'$} & 86/41 & 1.3066(19) & \\
            \textrm{$\tau'$} & 187/91 & 2.1765(13) & \\
            \textrm{$\gamma'+2\beta'$} $(= $d$\nu')$ & 2.3415 & 1.8811(19) & \\
            \hline
            \textrm{$\beta' - \beta/(\beta+1)$ } & & 0.00030(29) \\
            \textrm{$P_0' -$ Eqn. \ref{eq:prefactorRelation} } & & 0.00022(9) \\
            \textrm{$\gamma' - \gamma/(\beta+1)$} & & 0.0122(29) \\
            \textrm{$\tau' - \tau$} & & 0.0012(17) \\
        \end{tabular}
    \end{ruledtabular}
\end{table}

Assuming scaling relationships to hold \cite{Stauffer1985,Nakayama1994,Sahini1994}, and using that the fractal dimension of the lattice spanning cluster is identical for both protected and standard percolation ($d_f=d_f'$), we obtain the following relationships between the critical exponents for standard and protected percolation (as well as Eqn.~\ref{eq:betaRelation}):
\begin{equation}
    \gamma' = \frac{\gamma}{1+\beta},\qquad \tau' = \tau, \qquad \nu' = \frac{\nu}{1+\beta}
    \label{eq:protectedRelations}
\end{equation}
where $\nu$ is the critical exponent for the correlation length. We show in Fig.~\ref{fig:critExpFits} and Table \ref{tab:critexp} that these relationships are indeed borne out by our data and thus the scaling assumption appears to be justified for protected percolation as well; in particular, it is evident from Fig.~\ref{fig:critExpFits}c \ that $\tau' = \tau$ independent of any fitting procedure.  

The Harris criterion (Eqn.~\ref{eq:harrisCriterion}) is violated in three-dimensions for protected percolation, but not in two-dimensions. We list the value of the Harris criterion for each case in Table~\ref{tab:critexp}, demonstrating that it is violated for three-dimensional protected percolation ($1.876<2$). As such, any three-dimensional protected percolation system should display unique critical exponents that depend on the details of the impurities in each system. 

Our results are relevant to heavily-doped quantum critical systems such as Ce(Fe$_{0.76}$Ru$_{0.24}$)$_2$Ge$_2$~\cite{Montfrooij2019} and UCu$_4$Pd \cite{Aronson1995,Aronson2001}. Upon cooling, surviving clusters in Ce(Fe$_{0.76}$Ru$_{0.24}$)$_2$Ge$_2$ appear as short-range magnetic order; these clusters span identical numbers of moments along disparate crystallographic directions and they persist down to the lowest temperatures \cite{Montfrooij2007}. Given that the order is short-ranged and that the magnetic Ce-ions are separated by much larger distances along the c-axis than along the a-axis in this tetragonal compound ($c/a=2.5$), one would not have expected identical numbers of moments to be correlated along all directions. After all, the strength of the ordering interaction depends on the separation between Ce-ions. However, fully ordered, randomly formed clusters do come with the prediction of identical numbers of moments ordering, independent of the interaction strength \cite{Montfrooij2007}. In this system, the distribution of interatomic separations resulting from Fe/Ru doping resulted in a distribution of Kondo screening temperatures as these are exponentially sensitive to interatomic separations \cite{Sachdev2011,Endstra1993}.Therefore, each of the Ce-ions will be Kondo shielded at a unique temperature, thereby creating a protected percolation network upon cooling. 

Given that the Harris criterion is violated, this would offer natural explanation as to why universal critical exponents for quantum phase transitions have not been obtained despite such systems having been widely studied \cite{Stewart2001,*Stewart2006} If, in fact, heavily-doped quantum critical systems follow a three dimensional protected percolation model that violates the Harris criterion, we would not expect universal behavior because of the intrinsic disorder in these systems. We are currently performing MC simulations with added impurities in order to assess what level of impurity would lead to an experimentally observable change in critical behavior. 

In conclusion, protected percolation represents a new universality class with critical exponents that analytically relate to those of standard percolation \cite{Heitmann2014} and that violate the Harris criterion in three-dimensions. As such, impurities lead to system dependent critical exponents causing universal behavior to disappear. Protected percolation models heavily-doped quantum critical systems whose isolated magnetic clusters become ``protected" from Kondo screening. We have found quantum critical compounds that follow this model, such as Ce(Fe$_{0.76}$Ru$_{0.24}$)$_2$Ge$_2$~\cite{Montfrooij2019}. Our work leads to a natural explanation as to why universal critical exponents have not been found for quantum critical systems.

\bibliography{Percolation}

\end{document}